\documentclass[aps,prl,showpacs,preprintnumbers,twocolumn,superscriptaddress]{revtex4-2}
\usepackage{amsmath,amssymb}
\usepackage{bm}
\usepackage{tipa}
\usepackage{upgreek}
\usepackage{comment}
\usepackage{mathrsfs}
\usepackage{graphicx}
\usepackage{braket}
\usepackage{enumitem}
\usepackage{mathbbol}
\usepackage{booktabs}
\usepackage{gensymb}
\usepackage[normalem]{ulem}
\usepackage{color}
\usepackage[colorlinks,bookmarks=true,citecolor=blue,linkcolor=red,urlcolor=blue]{hyperref}
\usepackage{hyperref}

\usepackage{pifont}

\begin{document}

	\title{Variational wavefunctions for fractional topological insulators}

    \author{Glenn Wagner}
    \affiliation{Institute for Theoretical Physics, ETH Z{\"u}rich, 8093 Z{\"u}rich, Switzerland}
    \author{Titus Neupert}
    \affiliation{Department of Physics, University of Z{\"u}rich, Winterthurerstrasse 190, 8057 Z{\"u}rich, Switzerland}

	\begin{abstract}
    Twisted transition metal dichalcogenides (TMDs) host bands with opposite Chern number for the two spin species and could thus be host for fractional topological insulator states. In multicomponent quantum Hall systems, where the spins have equal Chern number, the resulting topological liquid states can be well modeled by trial wavefunctions such as the Halperin $(lmn)$ wavefunctions. These wavefunctions are exact zero energy states of certain short-range (pseudopotential) Hamiltonians. However, we show that such a construction fails in the case where the Chern numbers of the two spins are opposite, since the electrons with opposite Chern number cannot avoid one another. This underlines the importance of suppressing the short-range interspin Coulomb repulsion in order to realize fractional topological insulators in twisted TMDs. We introduce a trial wavefunction for the opposite Chern number Landau levels for filling factor $\nu=\frac{1}{3}+\frac{1}{3}$ in terms of pairing of composite fermions and show that it has good overlaps with the exact diagonalization ground state when the short-range Coulomb repulsion is sufficiently softened.
	\end{abstract}
	
 	\maketitle

\textit{Introduction}.---Multicomponent quantum Hall systems have a long history \cite{girvin1995multicomponentquantumhallsystems}. The usual problem set up consists of several flavours of electrons in Chern bands with the same Chern number $C$. The electron flavour could correspond to the physical spin, e.g. in GaAs heterostructures \cite{Eom2000}. Or it could correspond to a layer degree of freedom, such as in quantum Hall bilayers \cite{QHB_review}. Or it could be a valley degree of freedom such as in graphene or other systems \cite{Parameswaran_2019}. Multicomponent quantum Hall systems give rise to phenomena such as quantum Hall ferromagnetism \cite{Moon1995,Ezawa_2009}, skyrmions \cite{Sondhi1993} and non-Abelian quantum Hall states \cite{Barkeshli2010}. What these systems have in common is that the Chern number of all the electron flavours is the same. This is the physically relevant case for an externally applied magnetic field, since the sign of the Chern number is fixed by the sign of the magnetic field and in all the cases above, the electrons are located close together and hence the sign of the magnetic field must be the same. 

However, recent advances in twisted TMD heterostructures have enabled access to a very different situation: These materials can have Chern bands in the absence of an externally applied magnetic field. Since the bandstructure preserves time-reversal symmetry, the electron flavours must have opposite Chern number. In materials such as twisted MoTe$_2$ the spin-orbit coupling leads to spin-valley locking. For small twist angles, the band closest to charge neutrality is a flat Chern band and the bands in the opposite valleys have an opposite Chern number \cite{Wu2019}. The flat Chern band is an ideal setting for realizing a fractional Chern insulator \cite{neupert,sheng,regnault}. Indeed, experiments see signatures of spin-polarized fractional Chern insulators \cite{cai2023signatures,zeng2023integer,Park2023,Xu2023,park2024ferromagnetism,xu2024interplay} that spontaneously break time-reversal symmetry. On the other hand, fractional topological insulators are topologically ordered states which preserve time-reversal symmetry \cite{bernevigQSHE2006,Levin_Stern,Neupert2011FTI,Stern2015review,Neupert_2015,Levin2012classification}. Potential signatures of a fractional topological insulator have been observed at filling factor $\nu=-3$ in twisted MoTe$_2$ \cite{kang2024evidence_nature}, in which case electrons with opposite Chern number participate in the interacting state. However, a recent experiment suggests that time-reversal symmetry is broken in that state \cite{kang2025timereversalsymmetrybreakingfractional}, casting doubt on its interpretation as a fractional topological insulator. The set up of a multicomponent quantum Hall fluid with opposite Chern number Landau levels has not been experimentally relevant until now, hence it was not as extensively studied besides a few early works \cite{Chen2012FQHtorusFTI,furukawa2014global,Repellin2014FTI}. Motivated by the relevance for moiré materials, recent work has started looking at this set-up \cite{Mukherjee2019FQHsphereFTI,Bultinck2020mechanism,zhang2018composite,kwan2021exciton,kwan2022hierarchy,eugenio2020DMRG,stefanidis2020excitonic,chatterjee2022dmrg,myersonjain2023conjugate,yang2023phase,Wu2024,shi2024excitonic,kwan2024textured,abouelkomsan2023band,kwan2024abelianfractionaltopologicalinsulators}, however many interesting questions remain.

One place where the difference between the equal and opposite Chern number cases becomes clear is for the quantum Hall ferromagnet. Let us consider the case where electrons of both spins are at half-filling ie $\nu_\uparrow=\nu_\downarrow=\frac{1}{2}$ and they are interacting with a repulsive spin-isotropic Coulomb interaction. In the case $C_\uparrow=C_\downarrow=1$, the ground state is a quantum Hall ferromagnet, with in-plane spin polarization. This is the so-called Halperin (111) state \cite{Halperin:1983zz}, a generalization of which we will introduce below. The expectation value of the onsite repulsive interaction $V_0$ vanishes for this state. For the case $C_\uparrow=-C_\downarrow=1$, the band topology implies that the hybridization between the spins must vanish somewhere in the Brillouin zone. The in-plane quantum Hall ferromagnet is then frustrated \cite{Bultinck2020mechanism} and the resulting state, which must have out-of-plane spin polarization somewhere in the Brillouin zone, has been dubbed a Chern-textured insulator \cite{kwan2024textured,wang2024cherntexturedexcitoninsulatorsvalley}. This state has a non-zero expectation value for the onsite repulsive $V_0$ (see below).

The fact that $V_0$ has nonzero expectation value in the opposite Chern number case implies that the electrons in the two spins cannot avoid one another. This makes sense intuitively: In a magnetic field electrons undergo cyclotron motion and the Chern number determines the chirality of that cyclotron motion. With equal Chern numbers, the electrons are all circulating in the same direction and can avoid each other, whereas for opposite Chern numbers, they are circulating in opposite directions and cannot avoid each other and necessarily collide. This is particularly visible along the edge: For the chiral edge of a quantum Hall state there is no backscattering \cite{Buttiker1988}, whereas for the helical edge state of a quantum spin Hall state with counterpropagating modes there is backscattering \cite{Konig2013}. 

Finally, there is recent numerical evidence that shows that electrons in opposite Chern bands cannot avoid one another. The criteria for finding a fractional topological insulator as the ground state of opposite Chern number Landau levels has been studied in the context of twisted MoTe$_2$. Exact diagonalization studies show that the repulsive onsite $V_0$ must be suppressed in order to obtain a fractional topological insulator for a spin-isotropic interaction \cite{kwan2024abelianfractionaltopologicalinsulators}. Such a suppression of $V_0$ may come from phonons, Landau level mixing or dielectric engineering. The fractional topological insulator requires an ideal value of around $V_0/V_1\sim 1.2$, where $V_1$ is the first Haldane pseudopotential (see below). 

\textit{Failure of constructing repulsive wavefunctions}.---This difficulty of electrons avoiding each other can also be seen in terms of the wavefunctions that have been put forward to describe electrons in the opposite Chern number Landau levels. In particular, in the context of the quantum spin Hall effect, Ref.~\cite{Bernevig2006} proposes a wavefunction in the disk geometry of the form 
\begin{equation}
    \Psi_\textrm{QSH}=\prod_{i<j}(z_i-z_j)^l\prod_{i<j}(w_i-w_j)^{*m}\prod_{i,j}(z_i-w_j^*)^n,
\end{equation}
where $z_i=x_i+iy_i$ is the complex coordinate of the $i$th electron for the up spins and $w_i$ is the complex coordinate of the $i$th electron for the down spins, and $l,m,n$ are positive integers. The opposite Chern numbers for the two spins imply the wavefunction must be a holomorphic function of $z_i$ and an anti-holomorphic function of $w_i$.  Here, and in the following, for ease of notation we suppress the exponential factors associated with the Landau orbitals in the disk geometry. There is an issue with this wavefunction, in that the last term describes repulsion of the electrons in the $x$-direction, but attraction in the $y$-direction, which is unphysical. This state is shown to be energetically unfavourable and have poor overlaps with the exact diagonalization ground state \cite{Mukherjee2019FQHsphereFTI}.

We want to attempt to construct an analogue of the Halperin wavefunction for electrons in opposite Chern bands. For equal Chern bands, the Halperin $(lmn)$ wavefunction \cite{Halperin:1983zz} written in the disk geometry is
\begin{equation}
    \Psi_{lmn}=\prod_{i<j}(z_i-z_j)^l\prod_{i<j}(w_i-w_j)^m\prod_{i,j}(z_i-w_j)^n.
    \label{eq:Halperin_lmn}
\end{equation}
The fact that the electrons all live in a Landau level with Chern number $C=+1$ is reflected in the fact that the many-body wavefunction is a holomorphic function of all the coordinates. In this wavefunction, the electrons with up (down) spin have relative angular momentum $l$ ($m$). Electrons with opposite spin have relative angular momentum $n$. The Halperin $(lmn)$ wavefunction has filling factors $\nu_\uparrow=\frac{1}{l+n}$ and $\nu_\downarrow=\frac{1}{m+n}$. This wavefunction can be understood as the maximum density ground state of a Hamiltonian consisting of certain Haldane pseudopotentials \cite{Read1996}. The Haldane pseudopotential $V_k$ \cite{Haldane1983} leads to an energy penalty for electrons with relative angular momentum $k$. In real space, the interaction potential from the Haldane pseudopotential $V_k$ looks like $V(z)\propto L_k(-\ell_B^2\nabla^2)\delta(z)$, where $L_k$ is the $k$th Laguerre polynomial and $\ell_B$ is the magnetic length \cite{harper2015hofstadter}. We will consider an interspin interaction with non-zero $V_0,\dots,V_{n-1}$. Let us focus on the equal Chern number case first. We want to create a two-body wavefunction $f(z,w)$ such that $f(z,z)=0$. Writing the Taylor expansion
\begin{equation}
    f(z,w)=\sum_{mn}a_{mn}z^mw^n,
\end{equation}
the condition $f(z,z)=0$ has the solution
\begin{equation}
    f(z,w)=\sum_{m}a_{m0}(z-w)^m
\end{equation}
and requiring that the state has zero energy in the presence of non-zero $V_0,\dots,V_{n-1}$, as well as that it be the highest density state, fixes $a_{m0}=\delta_{nm}$. Taking the product over all pairs of particles leads to the last term in Eq.~\eqref{eq:Halperin_lmn}. The intraspin terms can be obtained similarly.

\begin{figure}
    \centering
    \includegraphics[width=\linewidth]{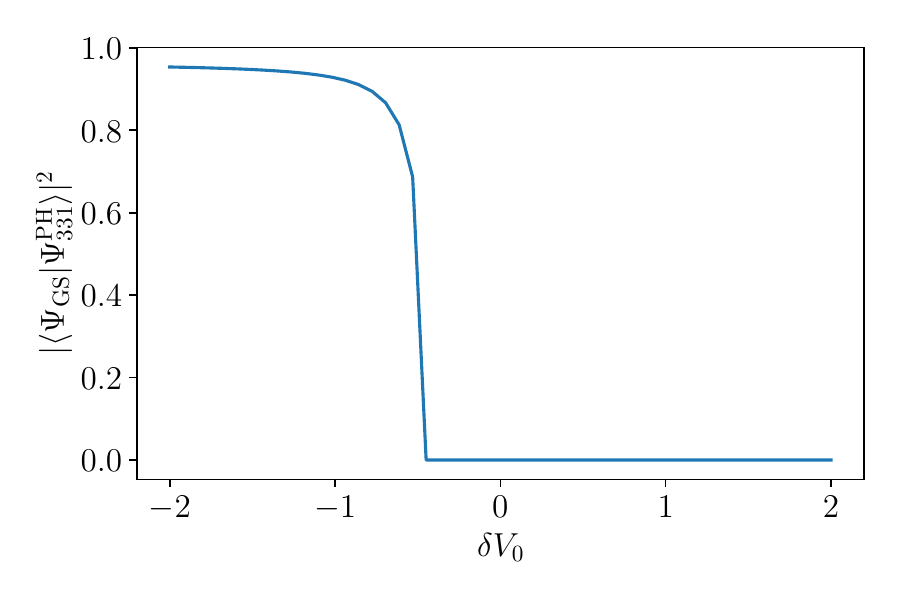}
    \caption{\textbf{Overlaps of the partially particle-hole transformed Halperin (331) state with the exact diagonalization ground state.} We perform exact diagonalization in the spherical geometry at $N=4+9$ electrons and $N_\phi=13$ flux quanta (this is the appropriate shift for the Halperin state on the sphere \cite{Papic2010}). The interaction is taken to be a spin-isotropic Coulomb interaction, where $\delta V_0$ has been added to the short-range Haldane pseudopotential $V_0$. The Coulomb interaction value is $V_0=0.95$. The partially particle-hole transformed Halperin (331) state $|\Psi_{331}^\mathrm{PH}\rangle$ only has good overlap with the exact diagonalization ground state $|\Psi_\mathrm{GS}\rangle$ once the interspin repulsion has been sufficiently softened.}
    \label{fig:Halperin_331_PH}
\end{figure}

Now let us consider the case with opposite Chern number. There, we require the wavefunction to be a holomorphic wavefunction of $z_i$ and an anti-holomorphic function of $w_i$. Let us write the Taylor expansion
\begin{equation}
    f(z,w^*)=\sum_{mn}a_{mn}z^mw^{*n},
\end{equation}
together with the condition enforced by $V_0>0$ which is $f(z,z^*)=0$. Due to the Cauchy-Riemann conditions, we can treat $z$ and $z^*$ as independent variables and hence 
\begin{equation}
    \frac{\partial^m}{\partial z^m}\frac{\partial^n}{\partial z^{*n}}f(z,z^*)=a_{mn}=0,
\end{equation}
showing that there is no nontrivial solution and the construction of the Halperin wavefunction fails. 

Reference~\onlinecite{Sodemann} proposes to express the Halperin wavefunction in a basis where one of the spins has been particle-hole transformed. For example, the partially particle-hole transformed Halperin (331) state, where the spin $\downarrow$ has been particle-hole transformed, describes the system at filling factor $\nu_\uparrow=\frac{1}{4}$, $\nu_\downarrow=\frac{3}{4}$. This state has been proposed for the time-reversal breaking state at $\nu=-3$ seen in Ref.~\onlinecite{kang2025timereversalsymmetrybreakingfractional}. In this formulation, before the particle-hole transformation, the Halperin wavefunction describes repulsion between electrons in the $C=+1$ band ($z_i$ in Eq.~\eqref{eq:Halperin_lmn}) and holes in the $C=-1$ band ($w_i$ in Eq.~\eqref{eq:Halperin_lmn}). After the particle-hole transformation, the wavefunction describes attraction between the opposite Chern number electrons. 

In Fig.~\ref{fig:Halperin_331_PH} we compute the overlap of the partially particle-hole transformed Halperin (331) state with the exact diagonalization ground state for a repulsive Coulomb interaction on the sphere. We vary the onsite Haldane pseudopotential $V_0$ relative to its value for the Coulomb interaction. Indeed, we see that this state has good overlap with the exact diagonalization ground state on the sphere, only once the interspin repulsion has been sufficiently reduced. This is consistent with the wavefunction having attractive correlations between opposite spin electrons. This highlights the difficulty of writing down a wavefunction describing the repulsion that we expect to be present.

So far we have based our argument on the failure of constructing repulsive wavefunctions for opposite Chern number electrons on a certain class of wavefunctions adapted from the Halperin wavefunctions. Another possible class of variational wavefunctions are Slater determinants. Certain Halperin $(lmn)$ wavefunctions such as the $(111)$ wavefunction are Slater determinants, but in general they are not. However, we can show that the onsite repulsion $V_0$ has non-zero expectation value for any Slater determinant ground state. The electron-hole condensate order parameter is $\Delta(\mathbf{r})=\langle c_{+,\mathbf{r}}^\dagger c_{-,\mathbf{r}}\rangle$, where $c_{\pm,\mathbf{r}}^\dagger$ is the creation operator at position $\mathbf{r}$ in the band with Chern number $C=\pm1$. In Ref.~\onlinecite{Bultinck2020mechanism} it was shown that this order parameter must vanish at least at one point in the unit cell, since the phase of the order parameter winds around the unit cell due to the opposite Chern number. Assuming a uniform charge density, this implies that at this point in the unit cell, both spins are occupied, however, they are also uncorrelated and hence cannot avoid one another. Therefore, $V_0$ must have a non-zero expectation value there. To show that, we compute the energy expectation value using the Hamiltonian
\begin{equation}
    H_0=V_0\int_\mathbf{r}c_{+,\mathbf{r}}^\dagger c_{+,\mathbf{r}}c_{-,\mathbf{r}}^\dagger c_{-,\mathbf{r}}
    \label{eq:S1}
\end{equation}
and using Wick's theorem as well as time-reversal symmetry to be
\begin{equation}
    \langle H_0\rangle=V_0\int_\mathbf{r}(\rho(\mathbf{r})^2-|\Delta(\mathbf{r})|^2),
\end{equation}
where the density is $\rho(\mathbf{r})=\langle c_{+,\bf{r}}^\dagger c_{+,\bf{r}}\rangle=\langle  c_{-,\bf{r}}^\dagger c_{-,\bf{r}}\rangle$. The integrand is positive semi-definite as can be seen from Eq.~\eqref{eq:S1}. At the same time, we know that $\Delta(\mathbf{r})=0$ at some point in the unit cell and assuming that $\rho(\mathbf{r})$ is uniform (or at least does not vanish at $\mathbf{r}$), we have that  $\langle H_0\rangle>0$. 

We conclude that the on-site repulsion leads to an energy penalty for a Slater determinant state, unless we break translational or time-reversal symmetry. This would imply a tendency for opposite Chern band systems to either spontaneously break time-reversal symmetry and form a spin polarized state, or to break translational symmetry and form a spin-density wave or a phase-separated state. This is indeed what is seen in exact diagonalization studies of opposite Chern bands \cite{Chen2012FQHtorusFTI,kwan2024abelianfractionaltopologicalinsulators,Mukherjee2019FQHsphereFTI}. This is also consistent with the experiments indicating that time-reversal symmetry is broken in the potential state exhibiting signatures of the fractional quantum spin Hall effect at $\nu=-3$ in twisted MoTe$_2$ \cite{kang2025timereversalsymmetrybreakingfractional}.

On the other hand, if the Chern numbers of the bands are equal, it would be possible to have $|\Delta(\mathbf{r})|>0$ everywhere and in particular there is an unobstructed quantum Hall ferromagnet with $|\Delta(\mathbf{r})|=\rho(\mathbf{r})$ such that $\langle H_0\rangle=0$.

\begin{figure}
    \centering
    \includegraphics[width=\linewidth]{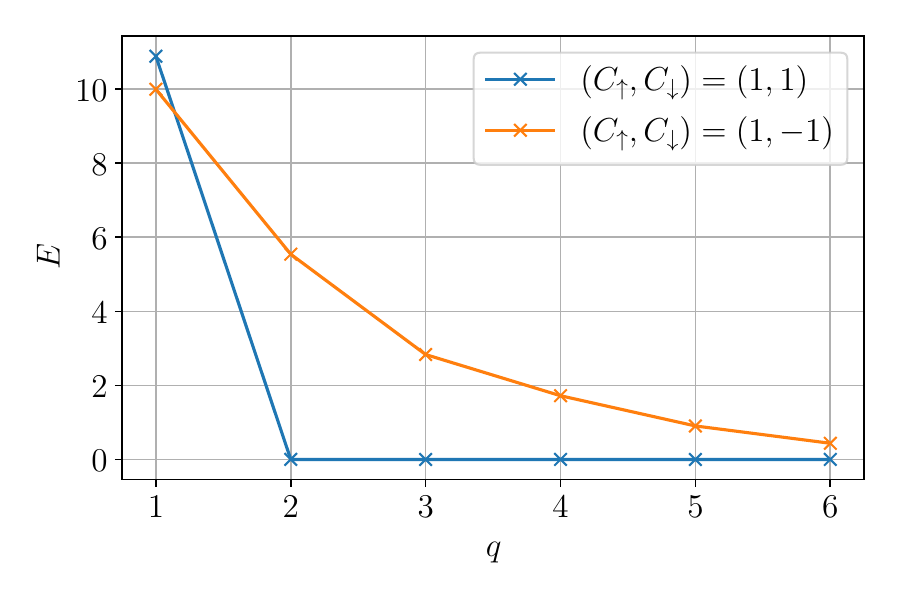}
    \caption{\textbf{Ground state energies for pseudopotential Hamiltonian.} We compute the ground state energy from exact diagonalization of the Hamiltonian with $V_0^{\uparrow\downarrow}=1$, $V_1^{\uparrow\uparrow}=V_1^{\downarrow\downarrow}=1$. We consider $N=6$ electrons (with $N_\uparrow=N_\downarrow=\frac{N}{2}$) in the spherical geometry with $N_\phi=q(N-1)-1$ flux quanta. The filling factor is $\nu_\uparrow+\nu_\downarrow=\frac{1}{2q}+\frac{1}{2q}$. We consider the cases where the Chern numbers of the spins are equal or opposite. For equal Chern numbers there is an exact zero energy state when $q\geq 2$, for $q=2$ this is the Halperin $(331)$ state. For opposite Chern numbers there is no such zero energy ground state. }
    \label{fig:kernel}
\end{figure}

\textit{Pseudopotential Hamiltonian}.---We construct a Hamiltonian that consists of $V_0^{\uparrow\downarrow}=1$, $V_1^{\uparrow\uparrow}=V_1^{\downarrow\downarrow}=1$ and all other pseudopotentials are set to zero and perform exact diagonalization on the sphere for a fixed number $N_\uparrow=N_\downarrow=\frac{N}{2}$ of electrons per spin species. We consider the system with a number of flux quanta $N_\phi=q(N-1)-1$, where $q$ is an integer. This corresponds to filling factor $\nu_\uparrow+\nu_\downarrow=\frac{1}{2q}+\frac{1}{2q}$ and is at the shift corresponding to the Halperin $(q+1,q+1,q-1)$ state \cite{Read1996}. In particular, for the case where the two spins have equal Chern number, the Halperin $(331)$ state will be an exact zero energy ground state for $q=2$ \cite{Read1996}. Therefore, in the equal Chern number case, the ground state energy vanishes for $q\geq2$. However, in the case where the Chern numbers are opposite, such a Halperin state cannot be written down as we argued above and hence there is no exact zero energy state at that filling. As shown in Fig.~\ref{fig:kernel}, the ground state energy for the opposite Chern number case is always positive, while in the equal Chern number case, the many-body Hamiltonian has a nontrivial kernel for $q\geq2$. It would be interesting to prove the absence of a kernel for the opposite Chern case rigorously. We conjecture an even stronger statement, namely that the ground state energy in the opposite Chern number case is always greater than the ground state energy in the equal Chern number case for any given filling due to $V_0^{\uparrow\downarrow}>0$.

\begin{figure}
    \centering
    \includegraphics[width=\linewidth]{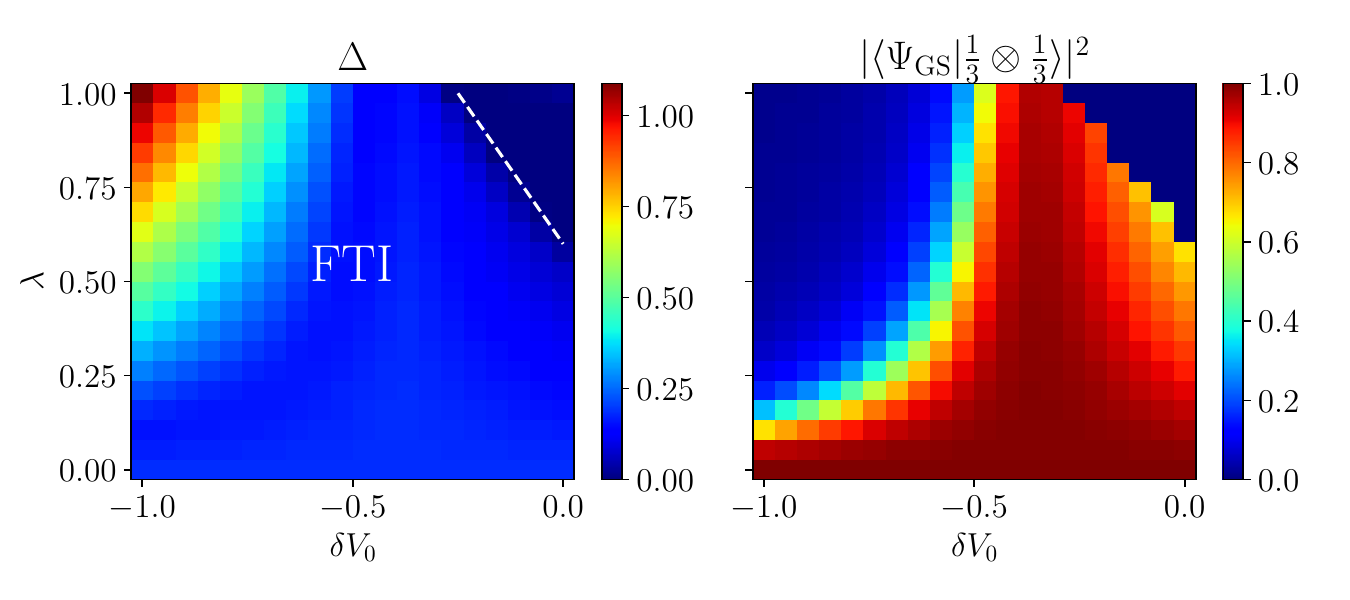}
    \caption{\textbf{Phase diagram for Coulomb interaction with opposite Chern Landau levels on the sphere.} We consider $N_\uparrow + N_\downarrow=5 + 5$ particles in $N_\phi=12$ flux quanta and vary the interspin suppression factor $\lambda$ and the onsite pseudopotential suppression $\delta V_0$. (left) Excitation gap $\Delta$. (right) Squared overlap of the exact diagonalization ground state $|\Psi_\mathrm{GS}\rangle$ with the decoupled Laughlin state $|\frac{1}{3}\rangle\otimes|\overline{\frac{1}{3}}\rangle$. There is a gap closing in the top left region of the phase diagram indicated by the white dashed line. In all the other parts of the phase diagram, the gap remains open and the state is adiabatically connected to the two decoupled Laughlin states, which is a fractional topological insulator (FTI) state. In the top left of the phase diagram, the overlap with the decoupled Laughlin state decreases, but the composite fermion trial wavefunction performs well, see Fig.~\ref{fig:overlaps}.}
    \label{fig:gap_10}
\end{figure}

\begin{figure}
    \centering
    \includegraphics[width=\linewidth]{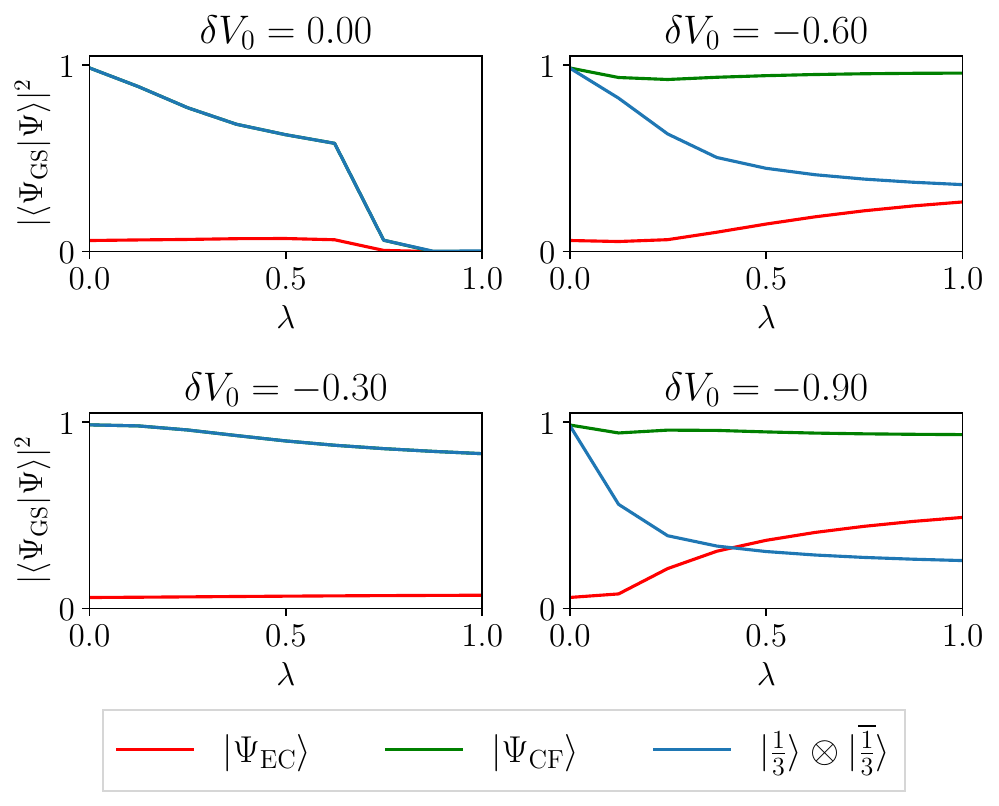}
    \caption{\textbf{Overlaps of three different wavefunctions with the exact diagonalization ground state as a function of interspin coupling $\lambda$.} We work in the spherical geometry for $N_\uparrow + N_\downarrow=6 + 6$ particles and $N_\phi=15$ flux quanta. We soften the onsite pseudopotential of the Coulomb interaction by $\delta V_0$. The three wavefunctions considered are (i) the exciton condensate $|\Psi_\textrm{EC}\rangle$, (ii) the optimized trial wavefunction for $s$-wave paired composite fermions as defined in Eq.~\eqref{eq:trialwfmaintext} $|\Psi_\mathrm{CF}\rangle$, (iii) two decoupled Laughlin states with opposite chirality $|\frac{1}{3}\rangle\otimes|\overline{\frac{1}{3}}\rangle$. The decoupled Laughlin state has high overlap with the ground state $|\Psi_\mathrm{GS}\rangle$ in the decoupled limit $\lambda=0$. The trial wavefunction $|\Psi_\mathrm{CF}\rangle$ performs well at large $\lambda$ only in the case where the onsite Coulomb repulsion is sufficiently suppressed. For $\delta V_0=0,-0.3$, the trial wavefunction does not lead to improved overlaps over the decoupled Laughin state (green curve lies under blue curve). }
    \label{fig:overlaps}
\end{figure}

\textit{Trial Wavefunctions}.---One of the great advances in understanding the quantum Hall effect came in terms of composite fermions, which consist of two flux quanta bound to an electron \cite{Jain_2007}. Motivated by the success of a class of composite fermion trial wavefunctions for quantum Hall bilayers, we will adapt these wavefunctions to the case of opposite Chern number Landau levels. In the quantum Hall bilayer case, one particle-hole transforms one of the layers and then uses a wavefunction that describes $s$-wave pairing of electron composite fermions in one layer with hole composite fermions in the other layer~\cite{Wagner2021}. In the case of opposite Chern number Landau levels, one forgoes particle hole transformation and directly pairs the electron composite fermions in both spins, bearing in mind that the flux attachment is opposite for opposite spins. Such a wavefunction can be written in the disk geometry as 
\begin{eqnarray}
    \Psi_\textrm{CF} &=& \mathcal{P}_\mathrm{LLL} \prod_{i<j}(z_i-z_j)^2(w_i-w_j)^{*2}\mathrm{det}(G)  \nonumber \\
    G(z_i, w_j)  &=& \sum_{n}g_n \phi_{n,m}(z_i)  \phi^*_{n,m}(w_j),
    \label{eq:trialwfmaintext}
\end{eqnarray}
where $g_n$ are the variational parameters of the trial wavefunction and $\phi_{n,m}$ are the single-particle Landau orbital wavefunctions in the $n$th Landau level with $L_z$ eigenvalue $m$. $\mathcal{P}_\mathrm{LLL}$ is the Jain-Kamilla projection to the lowest Landau level \cite{Kamilla}. In the following, we perform exact diagonalization in the sphere geometry (see Ref.~\onlinecite{Wagner2021} for details on how to write Eq.~\eqref{eq:trialwfmaintext} in the sphere geometry). For the interaction we use the repulsive Coulomb interaction $V_{\uparrow\uparrow}(r)=V_{\downarrow\downarrow}(r)=e^2/(4\pi\epsilon r)$ and $V_{\uparrow\downarrow}(r)=\lambda V_{\uparrow\uparrow}(r)$. Thus, $\lambda=1$ corresponds to the physically relevant limit of an interaction only dependent on the charge-density. We consider the system size $N_\uparrow=N_\downarrow=6$ with $N_\phi=15$ flux quanta (and $N_\uparrow=N_\downarrow=5$ with $N_\phi=12$) which, due to the shift \cite{Wen1992}, corresponds to filling factor $\nu=\nu_\uparrow+\nu_\downarrow=\frac{1}{3}+\frac{1}{3}$ on the sphere. We vary $V_0$ from the Coulomb value ($V_0=0.94$ for $N_\phi=15$) by some $\delta V_0$ in order to reduce the onsite repulsion.  We optimize the trial wavefunction to maximize the overlap with the exact diagonalization ground state. We use a single variational parameter $\alpha$ following Refs.~\onlinecite{Hu2024,Wagner2024} by using the ansatz $g_n=e^{\alpha n}$. Large $\alpha$ corresponds to tightly bound composite fermions. 

In Fig.~\ref{fig:gap_10}(left) we show the excitation gap in the $(\lambda,\delta V_0)$ phase diagram. For $(\lambda,\delta V_0)=(0,0)$ the two spins are decoupled and each at $\nu=\frac{1}{3}$ filling, they hence form two decoupled Laughlin states with opposite chirality. For $\delta V_0=0$, the gap closing at $\lambda\sim 0.7$ indicates that there is a transition to a state that is not a fractional topological insulator, this is consistent with Ref.~\onlinecite{kwan2024abelianfractionaltopologicalinsulators}. On the other hand, for sufficiently negative $\delta V_0$, there is no level crossing as a function of $\lambda$ and the system remains in the fractional topological insulator phase for all $\lambda\in[0,1]$.

In Fig.~\ref{fig:gap_10}(right) we also compute the overlap with the product of decoupled Laughlin states with opposite chirality for the two spins $|\frac{1}{3}\rangle\otimes|\overline{\frac{1}{3}}\rangle$, which has close to unity overlap with the ground state in the decoupled limit $\lambda=0$. If the composite fermions in each spin sector completely fill the $n=0$ composite fermion Landau level (``$\Lambda$-level"), they form an integer quantum Hall state and the trial wavefunction $\Psi_\textrm{CF}$ reduces to $|\frac{1}{3}\rangle\otimes|\overline{\frac{1}{3}}\rangle$. Thus, the decoupled limit is captured by our variational space (it corresponds to the limit $\alpha\to-\infty$). In Fig.~\ref{fig:gap_10}(right), we show that the decoupled Laughlin state only has good overlap with the ground state at $\lambda=1$ if $V_0$ is sufficiently suppressed.

Let us also introduce  the exciton condensate wavefunction $\Psi_\textrm{EC}=\textrm{det}[\sum_m \phi_{0,m}(z_i) \phi_{0,m}(w_j)^{*}]$, where the lowest Landau level wavefunctions in the disk geometry are $\phi_{0,m}(z)\propto z^m$. In the case where the spins have equal Chern number and $w_i^*$ represents spin $\downarrow$ holes, $\Psi_\textrm{EC}$ describes a quantum Hall ferromagnet.

In Fig.~\ref{fig:overlaps} we show the overlaps of the three wavefunctions $|\Psi_\textrm{EC}\rangle$, $|\Psi_\mathrm{CF}\rangle$, and $|\frac{1}{3}\rangle\otimes|\overline{\frac{1}{3}}\rangle$ with the ground state for four values of $\delta V_0$. The trial wavefunction $ \Psi_\textrm{CF}$ captures the exact diagonalization ground state well for all $\lambda$, but only once the onsite repulsion $V_0$ of the Coulomb interaction is sufficiently softened. This makes sense, since the trial wavefunction is not able to incorporate repulsive correlations between the opposite Chern band electrons. Crucially, the trial wavefunction captures the region for large negative $\delta V_0$, where the decoupled Laughlin states cease to be a good description. The exciton condensate performs better once the  onsite repulsion $V_0$ is small enough, since it describes attraction between opposite spin electrons. Since that wavefunction has no variational parameters, it performs significantly worse than the trial wavefunction. 

Finally, we note that extending these states to the lattice would require a construction similar to Ref.~\onlinecite{Wu2012}.

\textit{Conclusion}.---Recent experiments have motivated the study of multicomponent quantum Hall systems where the two components (which we have called ``spin” for specificity) have opposite Chern number. We have provided evidence that unlike in the equal Chern number case, here it is not possible to write down trial wavefunctions that are exact zero energy states of certain pseudopotential Hamiltonians. This is further supported by exact diagonalization numerics which show that in the opposite Chern number case, the kernel of the interaction Hamiltonian is empty. We write down a class of trial wavefunctions in terms of paired composite fermions which do capture the ground state of the system, albeit only in the case where the short-range repulsion is sufficiently softened. 

This work suggests that any system where on-site repulsion $V_0$ is present may have a difficulty forming a time-reversal symmetric and spin unpolarized fractional topological insulator and other competing states such as spin-polarized or phase-separated states may be the ground states instead. Our work highlights the importance of considering the energetics of the candidate states written down for fractional topological insulators. Furthermore, it shows the importance of considering effects such as band-mixing in the realistic models of twisted MoTe$_2$, since this has been shown to effectively reduce $V_0$ \cite{kwan2024abelianfractionaltopologicalinsulators,abouelkomsan2024nonabelianspinhallinsulator}. It would be interesting to test the states such as the partially particle-hole transformed Halperin state of Ref.~\onlinecite{Sodemann} or the non-Abelian fractional topological insulator of Ref.~\onlinecite{abouelkomsan2024nonabelianspinhallinsulator} in the realistic model of twisted MoTe$_2$, since these have been proposed to explain the experiments of Refs.~\onlinecite{kang2025timereversalsymmetrybreakingfractional,kang2024evidence_nature}. It is likely that including a mechanism to suppress $V_0$ such as band mixing will be necessary in the realistic model in order to stabilize these states.

There is also an alternative interpretation of the results. One can also consider particle-hole transforming one of the spins. In this particle-hole transformed basis the repulsive $V_0$ becomes an attractive one and the particles (meaning electrons in one spin and holes in the other) now have the same Chern number. The attractive $V_0$ wants to mediate attractive pairing between the particles, but this is obstructed due to the Chern number: It is known that one cannot have a uniform $s$-wave superconductor in a Chern band \cite{Murakami2003,Li2018}. Thus the $V_0$ term cannot be perfectly satisfied, electrons and holes in opposite spins cannot pair up completely, and hence electrons and electrons in opposite spins cannot avoid one another.

For the equal Chern number case where the interaction is given in terms of pseudopotentials, one can construct exact highest density zero-energy states. It would be interesting to place a lower bound on the energy for the same interaction and for the same filling, but in the opposite Chern number case. This may be possible using the recently developed quantum Hall bootstrap formalism which allows the computation of lower bounds on ground state energies \cite{gao2024bootstrappingquantumhallproblem}.

\begin{acknowledgements}
\textit{Acknowledgements}.--- G.W. would like to thank Inti Sodemann Villadiego for helpful discussions. Exact diagonalization calculations were performed using DiagHam. G.W.~is supported by the Swiss National Science Foundation (SNSF) via Ambizione grant number PZ00P2-216183. T.N. acknowledges support from the Swiss National Science Foundation through a Consolidator Grant (iTQC, TMCG-2\_213805). 
\end{acknowledgements}

\bibliography{bib}

\newpage
\clearpage
\begin{appendix}
\onecolumngrid
	\begin{center}
		\textbf{\large --- Supplementary Material ---\\Variational wavefunctions for fractional topological insulators}\\
		\medskip
		\text{Glenn Wagner and Titus Neupert}
	\end{center}
	
		\setcounter{equation}{0}
	\setcounter{figure}{0}
	\setcounter{table}{0}
	\setcounter{page}{1}
	\makeatletter
	\renewcommand{\theequation}{S\arabic{equation}}
	\renewcommand{\thefigure}{S\arabic{figure}}
	\renewcommand{\bibnumfmt}[1]{[S#1]}

\section{Phase diagram on the torus}

\begin{figure}[h]
    \centering
    \includegraphics[width=\linewidth]{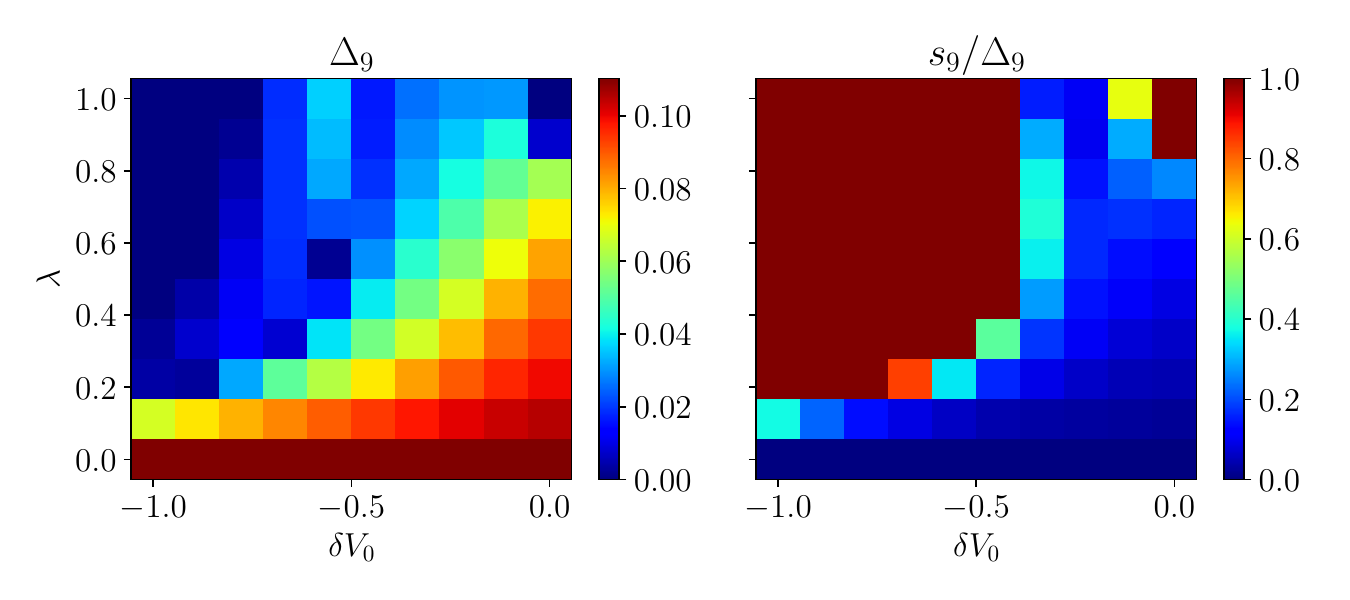}
    \caption{\textbf{Phase diagram on the torus.} We consider $N_\uparrow+N_\downarrow=5+5$ electrons in $N_\phi=15$ flux quanta. The interaction is taken to be the Coulomb interaction in the lowest Landau level truncated to include only the pseudopotentials $V_0,V_1,V_2,V_3$. We vary the interspin suppression factor $\lambda$ and the onsite repulsion suppression $\delta V_0$. (left) The excitation gap $\Delta_9$ between the ninefold degenerate ground state manifold and the excited states. (right) The maximum spread between the nine lowest energy states $s_9$ divided by the gap $\Delta_9$ between the ninth and tenth eigenvalue (see also Ref.~\onlinecite{kwan2024abelianfractionaltopologicalinsulators}). The fractional topological insulator at filling $\frac{1}{3}+\frac{1}{3}$ on the torus has a ninefold topological ground state degeneracy, so $s_9/\Delta_9$ quantifies how well this ground state degeneracy is satisfied. Small $s_9/\Delta_9$ indicates a good fractional topological insulator.}
    \label{fig:gap_10b}
\end{figure}

In Fig.~\ref{fig:gap_10b} we show the same phase diagram as in Fig.~\ref{fig:gap_10} in the main text, but for the torus geometry. The fractional topological insulator occupies a smaller region of the phase diagram on the torus as compared to the sphere. 

\end{appendix}

\end{document}